\documentclass[fleqn,usenatbib,letters]{mnras}

\usepackage{newtxtext,newtxmath}

\usepackage[T1]{fontenc}

\DeclareRobustCommand{\VAN}[3]{#2}
\let\VANthebibliography\thebibliography
\def\thebibliography{\DeclareRobustCommand{\VAN}[3]{##3}\VANthebibliography}

\usepackage{graphicx}	
\usepackage{amsmath}	
\usepackage{multirow}

\usepackage{scalerel,tikz}
\usetikzlibrary{svg.path}
\definecolor{orcidlogocol}{HTML}{A6CE39}
\tikzset{orcidlogo/.pic={
 \fill[orcidlogocol] svg{M256,128c0,70.7-57.3,128-128,128C57.3,256,0,198.7,0,128C0,57.3,57.3,0,128,0C198.7,0,256,57.3,256,128z};
 \fill[white] svg{M86.3,186.2H70.9V79.1h15.4v48.4V186.2z}
 svg{M108.9,79.1h41.6c39.6,0,57,28.3,57,53.6c0,27.5-21.5,53.6-56.8,53.6h-41.8V79.1z M124.3,172.4h24.5c34.9,0,42.9-26.5,42.9-39.7c0-21.5-13.7-39.7-43.7-39.7h-23.7V172.4z}
 svg{M88.7,56.8c0,5.5-4.5,10.1-10.1,10.1c-5.6,0-10.1-4.6-10.1-10.1c0-5.6,4.5-10.1,10.1-10.1C84.2,46.7,88.7,51.3,88.7,56.8z};
}}
\newcommand\orcidicon[1]{\href{https://orcid.org/#1}{\mbox{\scalerel*{
\begin{tikzpicture}[yscale=-1,transform shape]
\pic{orcidlogo};
\end{tikzpicture}
}{|}}}}

\title[Disrupted dwarf binary merger]{Disrupted dwarf binary merger as the possible origin of NGC 2419 and Sagittarius stream substructure}

\author[E. Y. Davies et al.]{
Elliot Y. Davies~\orcidicon{0000-0001-5996-4072}$^{1}$\thanks{E-mail: eyd20@cam.ac.uk},
Vasily Belokurov~\orcidicon{0000-0002-0038-9584}$^{1}$,
Stephanie Monty~\orcidicon{0000-0002-9225-5822}$^{1}$
and N. Wyn Evans~\orcidicon{0000-0002-5981-7360}$^{1}$.
\\
\\
$^{1}$Institute of Astronomy, University of Cambridge, Madingley Road, Cambridge CB3 0HA, UK}

\date{Accepted XXX. Received YYY; in original form ZZZ}

\pubyear{2015}

\begin{document}
\label{firstpage}
\pagerange{\pageref{firstpage}--\pageref{lastpage}}
\maketitle

\begin{abstract}
Using {\it N}-body simulations, we demonstrate that satellite dwarf galaxy pairs which undergo significant mixing ($\sim 6$ Gyr) can have their respective most bound particles separated great distances upon subsequently merging with a more massive host. 
This may provide an explanation as to the origin of the complex globular cluster NGC 2149, which is found within the tail of the Sagittarius dwarf spheroidal galaxy, yet separated from its central remnant by over 100 kpc. Dynamical investigations could support the chemical evidence which already points to the NGC 2419 being a nuclear star cluster.
Motivated by the distinct nature of NGC 2419, we run a suite of simulations whereby an initial {\it pre-infall} merger of two satellites is followed by a {\it post-infall} merger of the remnant into a MW-like host potential. 
We present a striking example from our suite in this work, in which this separation is reproduced by the most bound particles of the two pre-infall satellites. 
Additionally, this double merger scenario can induce unusual on-sky features in the tidal debris of the post-infall merger, such as clouds, overdensities, and potentially new arms that could resemble the bifurcation observed in Sagittarius.
\end{abstract}




\section{Introduction}

The Sagittarius  (Sgr) dwarf spheroidal galaxy is an ancient disrupting companion of the Milky Way (MW) whose vast sinuous tidal tails wrap around our Galaxy \citep[][]{ibata1995sgr,Majewski2003}. While it is estimated to have a present total mass around $4\times10^8$ M$_{\odot}$, work shows that Sgr will dissolve within the next Gyr \citep[][]{vasiliev2020last}, and was likely some orders of magnitude more massive in its past \citep[e.g.][]{niederste-ostholt2010re, bennett2022exploring}. The ongoing Sgr merger is the archetypal galactic cannibalism event, and stands as a testament to the ever-evolving nature of our home Galaxy, yet many aspects of the Sgr have not yet been explained -- for example, the bifurcation in its tidal stream \citep[][]{belokurov2006field,Koposov2012}. 


In this {\it Letter}, we consider a mechanism capable of producing noticeably distinct tidal debris sprays. In our model, two dwarf galaxies, with a mass ratio distinct from unity, coalesce and merge \textit{prior} to being accreted by the larger host, the Milky Way. In this scenario, the merged binary dwarf system has enough time to mix before the subsequent in-fall such that the two galaxies are not easily distinguishable in configuration space anymore. Even though they are now a part of the same satellite galaxy, stars from the two dwarfs have different orbital properties inside the merged binary system. Andromeda II is a prototype of such objects -- it has two populations with distinct kinematics as it is the remnant of a recent merger between two dwarf galaxies~\citep{Am14}. Having had a larger orbit around the common centre of mass, the lower-mass dwarf adds its material to that of the higher-mass one with a broader range of angular momenta and energy. This is similar to the \textit{cocoon} formation process proposed to explain the diffuse component of the stellar stream GD-1 \citep[][]{malhan2019butterfly,Carlberg2020}. 

The proposed binary dwarf galaxy scenario provides an attractive explanation for the existence of the peculiar globular cluster NGC 2419 found within the Sgr tidal debris. NGC 2419 is one of the largest and the most massive Galactic GCs, with the longest relaxation time \citep[][]{Baumgardt2018}. Together with $\omega$Cen and M54 (currently residing in the centre of the Sgr remnant), NGC 2419 has been suspected to be a nuclear star cluster from a disrupted dwarf galaxy \citep[see e.g.][]{Mackey2005, cohen2011peculiar, mucciarelli2012news, Pfeffer2021}. Although it has often been speculated that NGC 2419 may have been located in the Sgr itself~\citep[][]{Newberg2003,Ruhland2011,Precession,Sohn18}, no mechanism has so far been provided to generate its present position in the remote outskirts of the Galaxy or to rectify the existence of a secondary NSC within Sgr. NGC 2419 is located towards the Anticentre ($\ell =180.4^\circ, b=25.2^\circ$) at a Galactocentric distance of $r \sim$ 90 kpc ~\citep{harris2010new, baumgardt2021accurate}. This is far away from M54, which marks the centre of the Sgr dwarf, at $r \sim 24$ kpc and ($\ell = 5.6^\circ, b =-14.1^\circ)$. A merged binary system will have two nuclei, so the possibility of an explanation by this route is clear. What is missing is a mechanism through which the two nuclei can end up as widely separated as M54 and NGC 2419. We provide a simulation that answers this question unambiguously. 

In addition to providing an explanation for the Sgr-NGC~2419 system, a well-mixed binary-dwarf satellite could also permit the formation of a bifurcation in the tidal debris associated with pair post-infall into the MW.  To date, two hypotheses of the bifurcation seen in Sgr stream have been explored in the literature: i) that the two parts of the bifurcation result from two different wraps of the stream \citep[][]{Fellhauer2006} and ii) that the bifurcation is produced by sprays of stars from a rotating satellite's disk \citep[][]{penarrubia2010was}. The properties of each arm -- metallicities, distances, velocities \citep[][]{yanny2009tracing, niederste-ostholt2010re} -- are rather similar and thus do not look like two different wraps or two distinct un-mixed satellites \citep[but see][for evidence of slightly lower metallicity in the faint part of the leading bifurcation]{deBoer2015}. Little residual rotation is detected in the remnant's body today, leaving the hypothesis of the bifurcation resulting from the Sgr hosting a disc in the past disfavoured \citep[e.g. see][]{penarrubia2011no}. Recent analysis of the Sgr debris with the {\it Gaia} data has revealed that the bifurcation in both leading and trailing tails can be traced all the way back to the remnant \citep[][]{vasiliev2021tango,Ramos2022}, reaffirming that the stream bifurcation is associated with properties of the progenitor. Although our simulation is not tailored to reproduce the detailed Sgr stream bifurcation, it does possess features that suggest that a merged binary system is capable of resolving this riddle as well. 

The outline of this work is as follows. In Sec.~\ref{sec:methods} we explain the simulations methods, breaking our primary simulation into two distinct parts. We present our results and contextualise them in Sec.~\ref{sec:results}. Lastly, we conclude and summarise in Sec.~\ref{sec:summary}.

\begin{table}
\centering
\caption{Structural parameters of the two pre-infall merger satellites, and the single spheroid satellite.}
\label{tab:sat_params}
\resizebox{0.93\columnwidth}{!}{%
\begin{tabular}{|l|l|l|l|}
\hline
Satellite & Potential & Parameter [units] & Value \\ \hline\hline
\multirow{2}{*}{Major} & \multirow{2}{*}{Hernquist} & total mass {[}$10^{10}$ M$_{\odot}${]} & 3.0 \\
 &  & scale radius {[}kpc{]} & 12 \\ \hline
 \multirow{2}{*}{Minor} & \multirow{2}{*}{Hernquist} & total mass {[}$10^{10}$ M$_{\odot}${]} & 1.0 \\
 &  & scale radius {[}kpc{]} & 10 \\ \hline
 \multirow{2}{*}{Single spheroid} & \multirow{2}{*}{Hernquist} & total mass {[}$10^{10}$ M$_{\odot}${]} & 4.0 \\
 &  & scale radius {[}kpc{]} & 12.5 \\ \hline
\end{tabular}%
}
\end{table}

\section{Methods}\label{sec:methods}

In this work, the simulations are split into two distinct segments. We refer to the first simulation segment -- whereby two satellites merge in the absence of a host -- as the \textit{pre-infall} merger, and the second simulation segment -- whereby the resulting remnant is placed within the potential of a large MW-like host -- as the \textit{post-infall} merger. While a total suite of approximately 100 simulations were run to explore a wider parameter space of the two simulation segments, we focus on a single illustrative example which is a good representation of many in the sample. In these simulations, we varied the circularity of the less massive dwarf's orbit in the pre-infall merger, the duration of pre-infall merger, and the orientation of the resulting debris relative to the plane of the host. Throughout this work, we refer to the more massive and less satellites as the \textit{major} and \textit{minor} satellites respectively. We consider the ``stellar particles'' for both satellites to be the 10\% most bound (i.e. lowest energy) particles in their respective initial potentials.

\subsection{Pre-infall merger}

Both of the initial satellites consist only of a halo-like spheroidal \citet{hernquist1990analytical} potential
\begin{equation}
    \rho_{\rm sat}(r) = \frac{M}{2\pi} \frac{a}{r}\frac{1}{(r+a)^3},
\end{equation}
where $M$ is the total mass and $a$ is the scale radius. The positions and velocities of the satellites' particles are produced by sampling this potential, with an isotropic distribution function using the \textsc{GalaxyModel} tool in \textsc{Agama} \citep[][]{vasiliev2019agama}. The major satellite has a total mass of $M_{\rm maj} = 3\times10^5$ M$_{\odot}$ and consists of $N_{\rm maj} = 9\times10^5$ particles, whereas the minor satellite has a total mass of $M_{\rm min} = 1\times10^5$ M$_{\odot}$ and consists of $N_{\rm min} = 3\times10^5$ particles, ensuring the particles in both satellites have the same mass. The structural parameters of the two satellites can be found in Table~\ref{tab:sat_params}.

In the simulation we present here, the centre of the minor satellite is placed at the virial radius of the major satellite, with a circularity of $\eta = 0.6$, and the system is evolved in time using the python package \textsc{pyfalcON}, a stripped down python interface of the \textsc{gyrfalcON} code \citep[][]{dehnen2000celestial}. For this selected simulation set-up, we evolve the system for a ``mixing time'' of $t_{\rm mix} = 6$ Gyr, so that the two satellites are reasonably well mixed. By ``well mixed'' we simply refer to the pre-infall  merger having undergone around two merger timescales, as defined by Eq.~5 in \citet[]{boylan-kolchin2008dynamical}{}. We then save the positions and velocities of the merger for use in the post-infall merger simulation. Various snapshots of this process can be seen in Fig.~\ref{fig:preinfall_xy}, where we only show the stellar particles.

\subsection{Post-infall merger}

While the pre-infall satellite remnant remains represented as a collection of individually gravitating {\it N}-body particles, the host is taken to be the realistic MW-like fixed profile \textsc{MilkyWayPotential} from \textsc{Gala} \citep[][]{price-whelan2017gala}. The merger remnant of the two pre-infall satellites is placed 80 kpc above the centre of the plane of the host, with a circularity $\eta = 0.7$, to produce a moderately Sgr-like polar orbit with clearly elongated and wrapping tidal features. Note that despite aiming for a roughly Sgr-like orbit, we make no serious attempt to accurately model the orbit to match Sgr, nor to match the on-sky appearance of Sgr. We allow the debris to evolve for an ``infall time'' of $t_{\rm infall} = 5$ Gyr. In Fig.~\ref{fig:postinfall_xv}, we illustrate the 5 Gyr evolution of the post-infall merger, where we see numerous wrapping tidal tails in the stellar material.  Again the system is evolved using the \textsc{pyfalcON} code.

\subsection{Single spheroid merger}\label{sec:single_sph_method}

In addition to the satellite pair simulations discussed above, we also run a third simulation in which a single spheroid merges with the static host potential on the same orbit as the post-infall merger. The single spheroid is initialised from a Hernquist potential, with a total mass of $M_{\rm sph} = 4\times10^{10}$ M$_{\odot}$, a scale radius of 12.5, is made up of $12\times10^5$ particles with velocities sampled from an isotropic distribution function. Just as before, we allow this satellite to merge with the host for 5 Gyr, using the \textsc{pyfalcON} code. The stellar debris is again represented as the 10\% most bound particles. This single spheroid allows us to make comparisons of the well-mixed pre-infall merger debris with a more simple system made up of only one kinematic population.

\section{Results \& Discussion}\label{sec:results}

In this section we present the interesting features of our selected simulation, and briefly discuss likely mechanisms for their formation. We focus on the on-sky (right ascension \& declination) morphology, and the physical separation of the most bound particles of the respective satellite pairs.

\begin{figure}
    \centering
    \includegraphics[width=0.85\columnwidth]{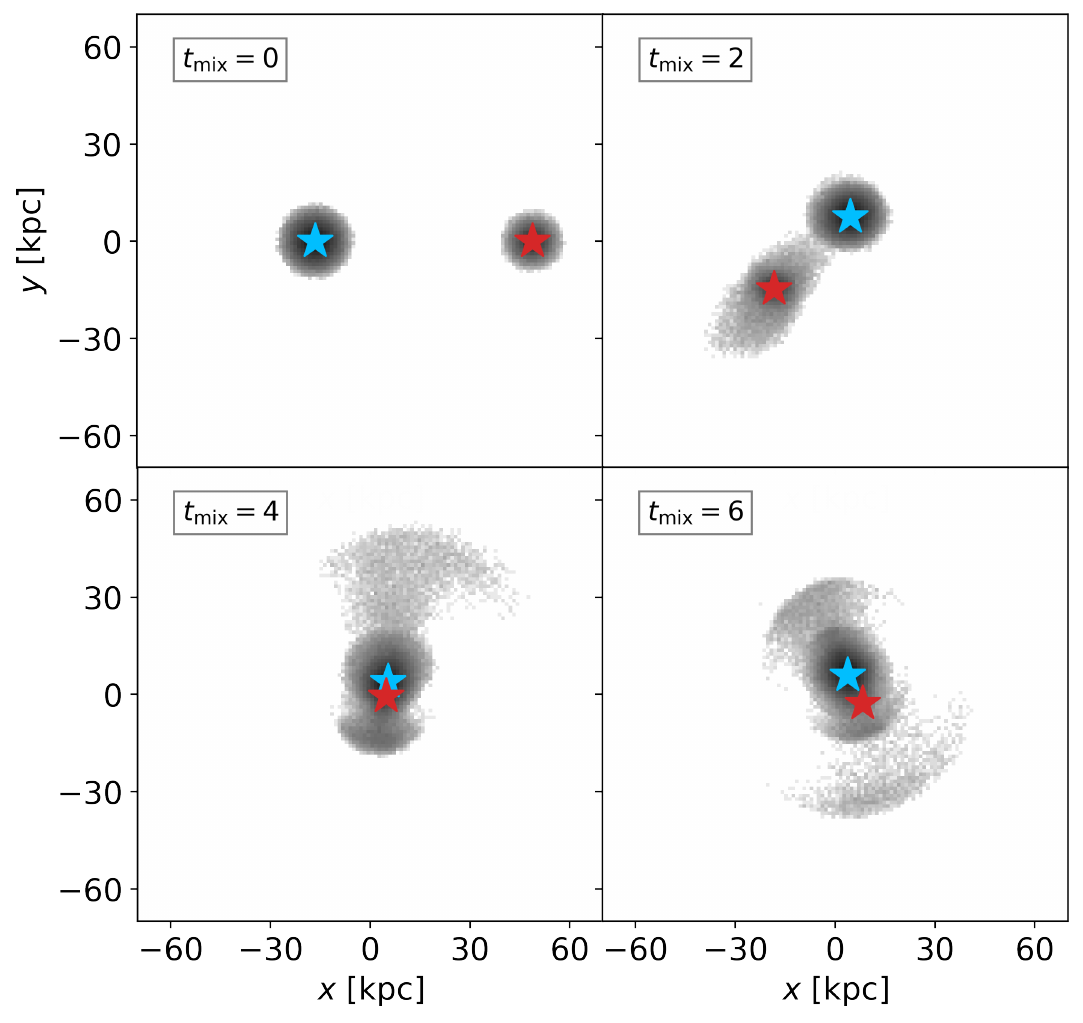}
    \caption{Evolution of the pre-infall merger in $(x,y)$ configuration space, showing snapshots at various pre-infall mixing times $t_{\rm mix}$. The dark 2-d histogram shows the log density of the stellar material for both the minor and major satellites. The red and blue coloured stars indicate the most bound particles of the minor and major satellites respectively. Features reminiscent of more well-mixed galaxies become apparent over time, such as the shells seen in the right-most panel. The units of $t_{\rm mix}$ are in Gyr.}
    \label{fig:preinfall_xy}
\end{figure}

\begin{figure*}
    \centering
    \includegraphics[width=0.85\textwidth]{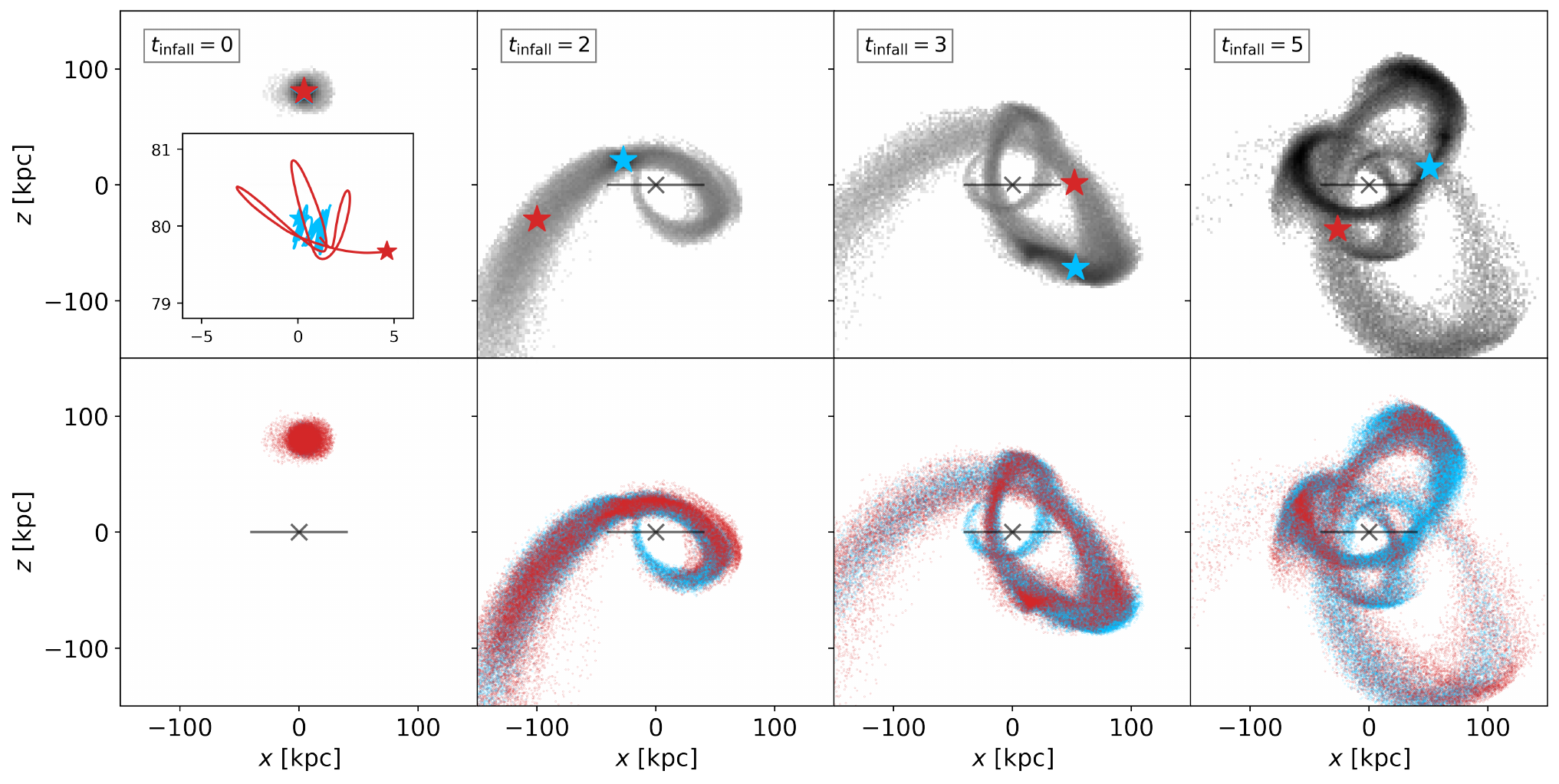}
    \caption{Evolution of the post-infall merger in $(x,z)$ configuration space, after allowing 6 Gyr of initial pre-infall mixing (as seen in the right-most panel of Fig.~\ref{fig:preinfall_xy}). \textit{Top:} the 2-d histogram shows the log density of the stellar material for both satellites, whereas the red and blue coloured stars indicate the most bound particles of the minor and major satellite respectively. Note the large separation of the most bound particles despite their initial close proximity. \textit{Inset panel:} Within the top leftmost panel, we show the orbits of the most bound particles for the 2 Gyr prior (and up to) the snapshot shown as $t_{\rm infall} = 0$. The red (blue) line shows the orbit of the minor (major) satellite's most bound particle. Note that the orbits of the minor satellite's most bound particle extend much further. \textit{Bottom:} scatter plot showing the two different mass satellites in different colours that match the most bound particle in the top row. Note that the streams are well-mixed and the orbits of particles from both satellites mostly overlap. The units of $t_{\rm infall}$ are in Gyr, and the faded black line is a visual aid to represent the $(x,z)$ plane of the host galaxy.}
    \label{fig:postinfall_xv}
\end{figure*}

\begin{figure}
    \centering
    \includegraphics[width=0.82\columnwidth]{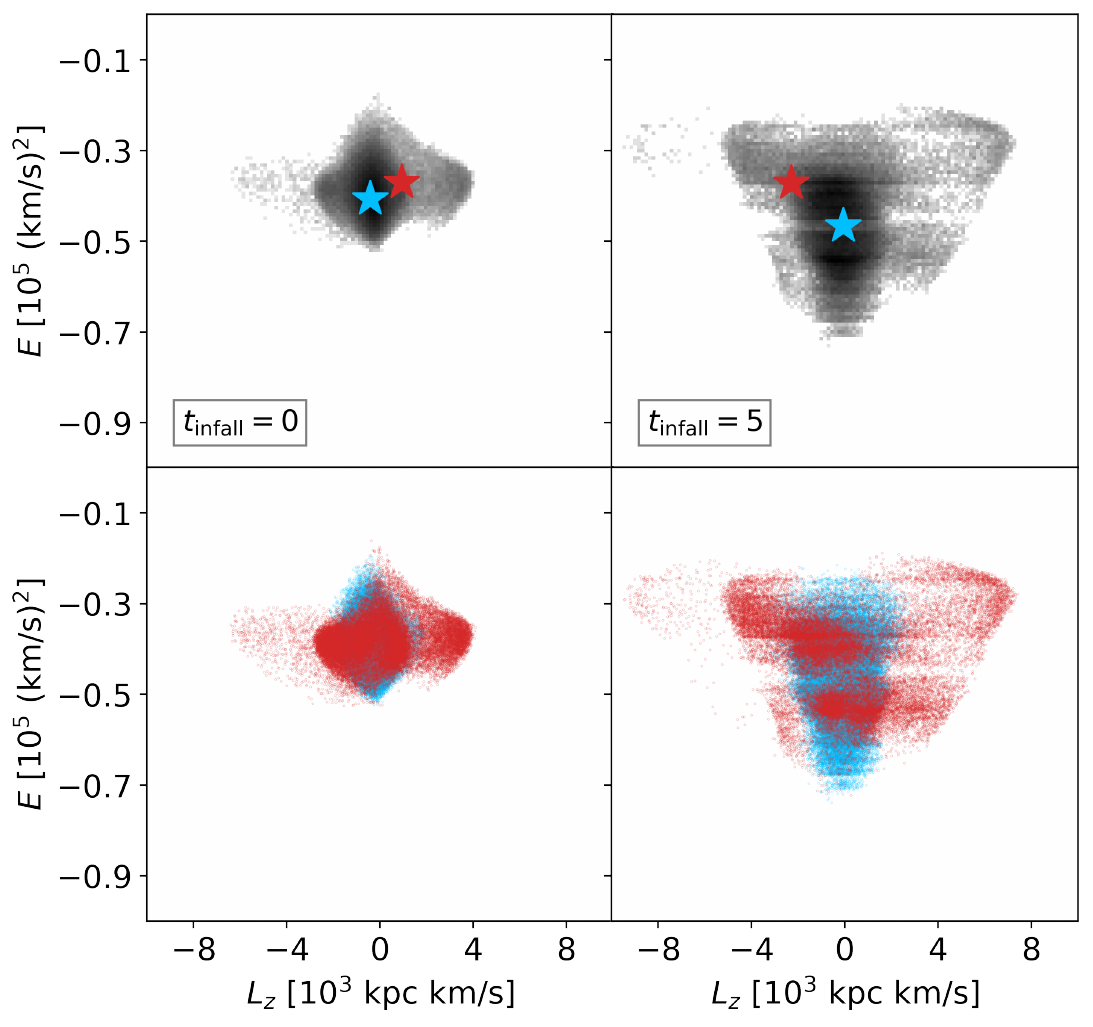}
    \caption{The first and last snapshot in $(L_z,E)$ space of the post-infall simulation, after allowing the initial pre-infall merger to mix for 6 Gyr. Therefore, the left and right columns of this figure correspond to the same snapshots as the left-most and right-most columns from Fig.~\ref{fig:postinfall_xv}. \textit{Top:} 2-d log density histogram of the stellar material for both satellites. \textit{Bottom:} scatter plot showing the two different mass satellites in different colours; the minor satellite is in red and the major satellite is in blue. Note how, while initially both satellites occupy a more narrow range of $E$ and $L_z$, the lighter satellite spreads out in both directions to form a high $|L_z|$ cloud around the more massive satellite.}
    \label{fig:postinfall_lze}
\end{figure}

\begin{figure*}
    \centering
    \includegraphics[width=0.82\textwidth]{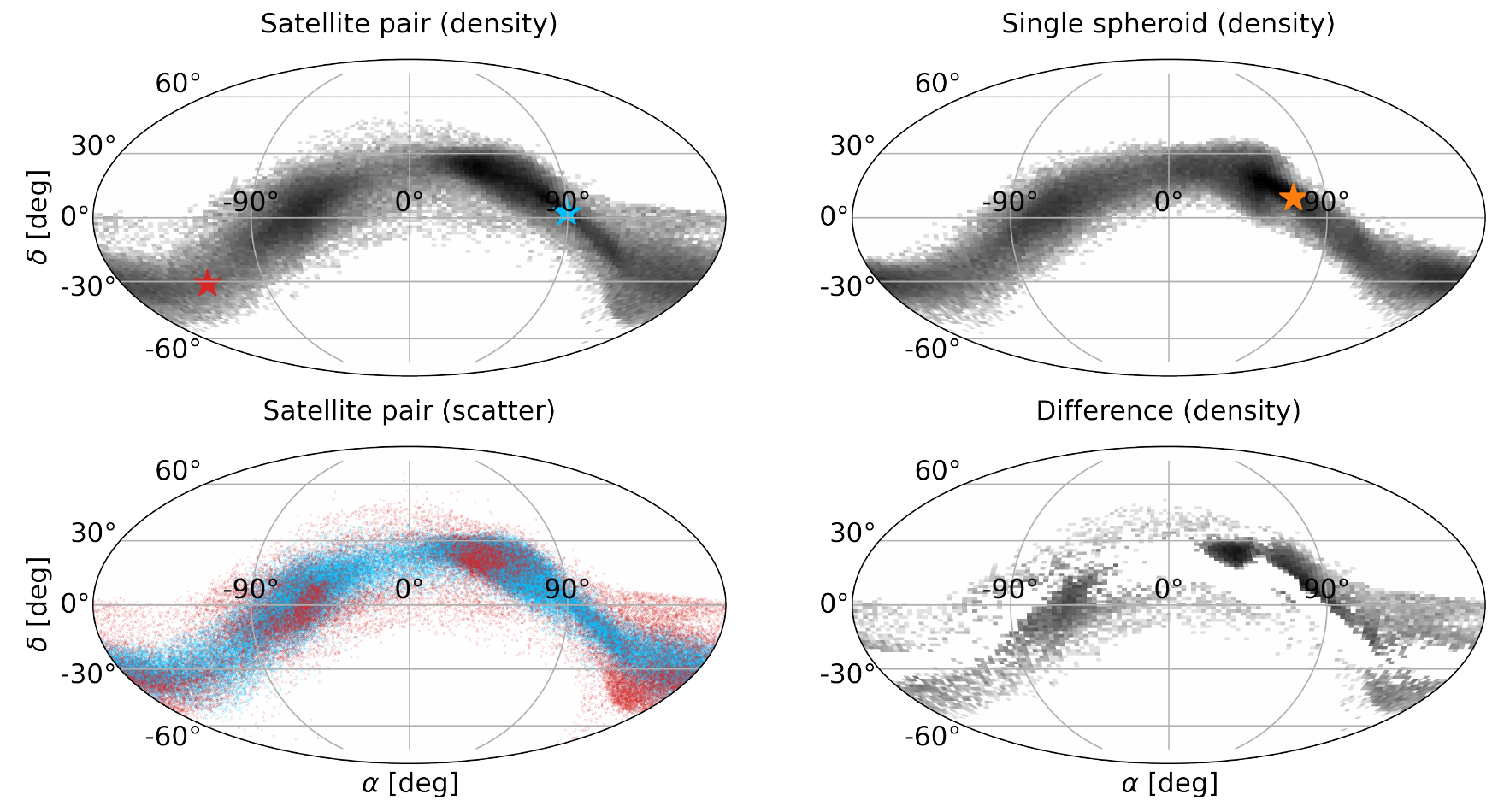}
    \caption{The on-sky coordinates of right ascension and declination $(\alpha, \delta)$, after an infall time of $t_{\rm infall} = 5$ Gyr. \textit{Top row:} 2-d log density histogram of the stellar material for both satellites on the left, and for a single Hernquist spheroid on the right. Superimposed are coloured stars which indicate the most bound particles of the minor (red), major (blue), and single spheroid (orange) satellite respectively. \textit{Bottom left:} scatter plot showing the different satellites in different colours; the minor satellite is in red and the major satellite is in blue. Note the large on-sky separation of the most bound particles, which re-illustrates the separation seen in Fig.~\ref{fig:postinfall_xv}. \textit{Bottom right:} 2-d log density histogram of the difference between the on-sky tracks of the satellite pair and the single spheroid; it is clear that additional structure has formed around the major satellite that is not present in the single spheroid. This is most striking at $\alpha > 90$ and $\delta \sim 0$. }
    \label{fig:onsky_compare}
\end{figure*}

\subsection{Separation of most bound particles}

To investigate whether the large separation of NGC 2419 and the central Sgr remnant could be achieved by the infall of a merger remnant of two satellites, we follow the trajectories of the most bound particles (MBPs) of each pre-infall satellites. These MBPs act as a proxy for the NSCs of their respective satellites. We aim to see whether, despite undergoing a merger for 6 Gyr, distinct kinematic populations can still be found within the post-infall debris. In Fig.~\ref{fig:preinfall_xy}, we present several snapshots in $(x,z)$ space of the pre-infall merger that was selected for this work, where we show only the stellar particles. The MBP of the major satellite is represented by a red star, whereas the MBP of the minor satellite is represented by a blue star. Fig.~\ref{fig:preinfall_xy} shows that after 6 Gyr, the two satellites have merged for a significant amount of time, such that typical long-timescale features have formed -- shells. It can be clearly seen that the location of the two MBPs are in the centre of their respective satellites in the initial snapshot. However, by the final snapshot, the orbit of the minor satellite's MBP is such that it (and other minor satellite particles) may be more prone to being tidally separated. 

Naively, we may anticipate that after a mixing time of 6 Gyr, the two satellites MBPs would have sunk to the bottom and would both be quite tightly bound to the centre of the pre-infall merger remnant. However, in Fig.~\ref{fig:postinfall_xv}, we show that the MBP of the minor satellite is ejected from the pre-infall remnant and, while staying within the tidal stream of the debris, is separated from the major satellite's MBP by approximately $110$ kpc. The timescale on which this happens varies depending on the initial merger set up, but can happen within a few Gyr, as evident by Fig.~\ref{fig:postinfall_xv}. This is reminiscent of the location of NGC 2419's  within the Sgr debris stream.

The ejection of the minor satellites' MBP relates to the particle's orbit, which we present in the inset panel within Fig~\ref{fig:postinfall_xv}. The more eccentric orbits of particles in the minor satellite, resulting from the nature of the pre-infall merger, make them more prone to being tidally stripped during the post-infall merger. If the minor satellite's MBP is close to apocentre as the pre-infall merger remnant falls into the host galaxy, then the separation between the two MBPs can be large. Among our other simulations in the suite, we find numerous examples of this circumstance being repeated, as well as a number of examples where the separation is much less dramatic. A reduced separation is simply due to the minor satellite's MBP being closer to its pericentre as the pre-infall merger remnant approaches the host. The inset panel of Fig~\ref{fig:postinfall_xv} nicely demonstrates that the minor satellite's MBP is more eccentric and so reaches farther out along its orbit. This is unlike the major satellite's MBP, which has a much more compact orbit, and doesn't extend to the same radii as the minor satellite's MBP. Despite the separation of the MBPs, the stream tracks from the two satellites mostly overlap, as shown by the coloured particles in the bottom row of Fig.~\ref{fig:postinfall_xv}.

As an additional investigation into the differing behaviour of the two satellites, we look to "integral of motion" space, specifically $(E,L_z)$, where the energy is calculated in the potential of the host. Though of course neither $E$ nor $L_z$ is conserved, this space is often used to study accretion onto the Milky Way as a proxy for action space~\citep[e.g.,][]{My18}. In Fig.~\ref{fig:postinfall_lze}, we show the first ($t_{\rm infall} = 0$ Gyr) and final ($t_{\rm infall} = 5$ Gyr) snapshots of the post-infall merger. We can see that in the final snapshot, the minor satellite debris has formed a higher $|L_z|$ cloud around the lower $|L_z|$ major satellite. Both satellites have their energy distributions widened significantly after falling into the host.

\subsection{Stream track appearance}

Can such a double merger scenario emulate the on-sky (right ascension $\alpha$ and declination $\delta$) appearance of the Sgr galaxy streams? We explore how well our scenario recreates the bifurcation in Sgr's tail \citep[e.g.][]{belokurov2006field}. Therefore, in addition to the $(x,z)$ space presentation of the stream track, we also plot the on-sky coordinates of the final snapshot in Fig.~\ref{fig:onsky_compare}. In this figure, we plot both the final snapshot of the pre-infall merger remnant and of the single spheroid merger described in Sec.~\ref{sec:single_sph_method}, for comparison. The distribution of the minor and major satellite debris are shown in red and blue colours. Again, coloured stars indicate the location of each MBP in the respective colours, with an orange marker for the single spheroid.

In Fig.~\ref{fig:onsky_compare} we see the separation of the MBPs reillustrated, as they are separated by about $144^\circ$ in $\alpha$ and $32^\circ$ in $\delta$ on the sky. In addition to this separation, we also see new on-sky features that are distinct to those found in the case of the single spheroid. 
To emphasise these features, we plot the difference between the satellite pair and the single spheroid in the bottom right panel of Fig.~\ref{fig:onsky_compare}.
The minor satellite particles create a cloud around the major satellite particles, in addition to forming new overdensities along (and around) the main on-sky track. The most notable difference is the additional ``arm'' of debris that can be seen clearly at $\alpha > 90$ and $\delta \sim 0$. While the exact morphology of the features is highly sensitive to the simulation set-up, these results demonstrate that our binary dwarf scenario is capable of creating a range of on-sky features, including a stream bifurcation.

\section{Summary}\label{sec:summary}

Motivated by the bifurcated tidal debris of the Sagittarius (Sgr) dwarf galaxy, we ran a suite of $N$-body simulations of binary dwarf galaxies merging with the Milky Way Galaxy. In this {\it Letter}, we present one simulation from our suite that illustrates the morphology of the debris when two coalescing satellites are engulfed by a larger host galaxy. We allow the the {\it pre-infall} merging satellite pair to mix for 6 Gyr, and then subsequently introduce a host galaxy as a static potential in which the remnant of the initial merger undergoes a second {\it post-infall} merger for 5 Gyr. The simulation shows a feature that is common in the suite: the separation of the most bound particles (MBP), acting as a proxy for the nuclear star clusters (NSC), of the two merging satellites after disruption by the host can be surprisingly large. 

By following their trajectories, we show that the MBP for both satellites can be separated by great distances ($\gtrsim 100$ kpc). In our simulation, this is caused by the ejection of the smaller satellites MBP, resulting from the more eccentric nature of its orbit after the pre-infall merger. 
The feature is reminiscent of the kinematic association between Sgr stream and the chemically complex, massive globular cluster NGC~2419. Like M54, which marks the centre of the Sgr remnant, NGC~2419 has often been considered a candidate nuclear star cluster. Our simulation shows that the merging of a binary dwarf galaxy can provide a natural explanation of the existence of two nuclear star clusters widely separated in the same tidal debris stream. The two satellites' MBPs separation is reinforced by viewing the simulation in on-sky coordinates, where the two particles are separated by over 140 degrees.

In addition, by comparison with a single spheroid merging with the host, we show that this satellite pair scenario can create distinct on-sky (right ascension \& declination) features that could emulate the Sgr bifurcation. Having two separate, yet well-mixed populations, allows for the creation of features that may not be possible given only one infalling satellite, such as additional overdensities and arm-like structures. Our large suite of simulations spans a wide parameter space. The morphology of the post-infall satellite pair debris depends on the orientation of the pre-infall debris relative to the host, the circularity of the pre-infall merger and the structural parameters of the dwarfs. We intend to follow up this illustration work with a statistical examination of the various morphologies that can be achieved. Moreover, since our mechanism predicts that the Sgr stream could be populated with debris from a second dwarf galaxy, we will also explore the possibility of detecting two distinct chemical populations within the Sgr stream debris.

\section*{Acknowledgements}

The authors would like to thank the rest of the Cambridge streams group, and specifically Adam Dillamore and GyuChul Myeong for useful discussion. We especially thank Eugene Vasiliev for the \textsc{AGAMA} software which was used throughout this work. EYD thanks the Science and Technology Facilities Council (STFC) for a PhD studentship (UKRI grant number 2605433). 

\section*{Data Availability}

The simulations in this project can be reproduced with publicly available software using the description provided in Sec.~\ref{sec:methods}.



\bibliographystyle{mnras}
\bibliography{distant} 




\appendix


\bsp	
\label{lastpage}
\end{document}